\documentclass[useAMS,usenatbib]{mn2e}
\usepackage{graphicx}

\title[The GRB luminosity function in the internal shock model confronted
to observations]
{The GRB luminosity function in the internal shock model confronted
to observations}
\author[H.~Zitouni~et~al.]{
H.~Zitouni$^{1,2}$\thanks{E-mail: \texttt{zitouni@iap.fr}~(HZ); \texttt{daigne@iap.fr}~(FD); \texttt{mochko@iap.fr}~(RM)}, 
F.~Daigne$^{1,3}$,
R.~Mochkovich$^{1}$,
and T.~H.~Zerguini$^{2}$\\
$^{1}$Institut d'Astrophysique de Paris, UMR7095 CNRS, Universit\'{e} Pierre \& Marie Curie-Paris 6, 
98 bis boulevard Arago, 75014 Paris, France.\\
$^{2}$Universit\'{e} des Sciences et Technologies Houari Boumebedienne,
BP 32 16123 Bab-Ezzouar Alger, Algeria.\\
$^{3}$Institut Universitaire de France.}

\begin{document}

\date{Accepted 2008 February 21. Received 2008 January 15; in original form 2007 October 26}
\volume{386} \pagerange{1597--1604} \pubyear{2008}
\maketitle
\label{firstpage}

\begin{abstract}
We compute the expected luminosity function of GRBs in the context
of the internal shock model. We assume that GRB central engines
generate relativistic outflows characterized by the respective 
distributions
of injected kinetic power $\dot E$ and contrast in Lorentz factor 
$\kappa=\Gamma_{\rm max}/\Gamma_{\rm min}$. We find that if the
distribution of contrast extends down to values close to unity (i.e.
if both highly variable and smooth outflows can exist) the luminosity 
function has two branches. At high luminosity it follows the
distribution of $\dot E$ while at low luminosity it is close to a
power law of slope $-0.5$.  
We then examine if existing data can constrain the luminosity function.  
Using the $\log{N}-\log{P}$ curve,
the $E_{\rm p}$ distribution of bright BATSE bursts and the XRF/GRB ratio
obtained by HETE2
we show that
single and broken power-laws can provide equally good fits of these
data. Present observations are therefore unable to favor one form of the
other.
However when a broken power-law is adopted they clearly 
indicate a low luminosity slope
$\simeq -0.6\pm 0.2$, compatible with the prediction of the internal shock model.

\end{abstract}

\section{Introduction}
The isotropic luminosity of long gamma-ray bursts is known
to cover a wide range from underluminous, nearby bursts such as
GRB 980425 or GRB 060218 (with $L\la 10^{47}$ erg.s$^{-1}$)
to ultrabright objects like GRB 990123 ($L\ga 10^{53}$ erg.s$^{-1}$).
While it has been suggested that the weakest GRBs could simply be normal 
events seen off-axis \citep{yamazaki:03},  this possibility has been recently discarded both
from limits on afterglow brightness and for statistical reasons \citep{soderberg:04,daigne:07}. The difference
of six orders of magnitude between the brightest and weakest GRBs is therefore 
probably real. The parameters (stellar rotation, metallicity, etc.) which
are responsible for this diversity in radiated power are not known.
However, in the restricted range $10^{51}\la L\la 10^{53}$ erg.s$^{-1}$ the 
value of the isotropic luminosity is possibly fixed by the opening angle
of the jet which may always carry the same characteristic energy \citep{frail:01}.  

The purpose of this paper is to see how 
basic theoretical
ideas 
and existing data
can be used to constrain the GRB luminosity function (hereafter LF)
$p(L)$. First, we should insist that $p(L)$ here represents the ``apparent"
LF which includes viewing angle effects and beaming statistics
(i.e. bursts with narrow jets are more likely seen off-axis and
therefore under-represented in the distribution). It is
therefore different from the ``intrinsic" LF $p_0(L)$ which would be obtained with 
all GRBs seen on-axis. In the lack of a complete, volume limited sample
of GRBs with known redshift, only indirect observational indicators such as
the $\log{N}-\log{P}$ plot can constrain the LF.
These indicators however depend not only on $p(L)$ but also on the GRB rate and
spectral energy distribution. The simplest possible form for $p(L)$ 
is a single power law $p(L)\propto L^{-\delta}$ between $L_{\rm min}$
and $L_{\rm max}$. Together with the parameters describing the GRB rate and
spectral shape, $\delta$, $L_{\rm min}$ and $L_{\rm max}$ can be 
adjusted to provide the best possible fit of the available indicators.
Considering the mixing of the LF with other quantities in the fitting 
process it is remarkable that studies using different observational
constraints have converged to a similar value of the slope $\delta\sim 1.5$
- $1.7$ \citep[e.g.][]{firmani:04,daigne:07}. 

In a second step one can consider the more general case of a broken power law LF with now five 
parameters: $L_{\rm min}$, $L_{\rm max}$, the two slopes $\delta_1$ and
$\delta_2$ and the break luminosity $L_{\rm b}$. We will see
in Sect.2 that there is some
indication that the internal shock model of GRBs can produce a broken 
power law LF and we want to check if it is also favored by the 
existing observational data. As in our previous study we have used 
a Monte Carlo method to generate a large number of synthetic events
where the parameters defining the burst properties are varied within 
fixed intervals. Preferred values of the parameters are those which yield
the minimum $\chi^2$ for a given set of observational constraints.
We summarize these constraints and present the Monte Carlo
simulations in Sect.3. We discuss our results in Sect.4 
and Sect.5 is our conclusion. 

\section{The GRB luminosity function in the internal shock model}
\label{sec:is}

The internal shock model \citep{rees:94} is the most 
discussed solution to explain the prompt gamma-ray emission in GRBs. 
In this section, we demonstrate that it naturally leads to a broken 
power-law LF with a low-luminosity slope close to $-0.5$.\\

\subsection{Internal shock efficiency and luminosity function}
In the context of the internal shock model \citep{rees:94}, 
the prompt gamma-ray emission is produced by relativistic electrons 
accelerated in internal shocks propagating within a relativistic outflow. The (isotropic equivalent) radiated luminosity $L$ is then a fraction of the (isotropic equivalent) 
rate of kinetic energy injected in the flow $\dot{E}$
\begin{equation}
L= f\left(\kappa\right) \dot{E}\ .
\end{equation}
The efficiency $f\left(\kappa\right)$ is the product of three terms
\begin{equation}
f\left(\kappa\right) \simeq f_\mathrm{rad}\ \epsilon_\mathrm{e}\ f_\mathrm{dyn}\left(\kappa\right)\ ,
\end{equation}
where ({\it i}) $f_\mathrm{dyn}\left(\kappa\right)$ is the fraction of the kinetic energy which is converted by internal shocks into internal energy (``dynamical efficiency''). This fraction depends mainly on the contrast $\kappa=\Gamma_\mathrm{max}/\Gamma_\mathrm{min}$ of the 
Lorentz factor distribution in the relativistic outflow;
({\it ii}) $\epsilon_\mathrm{e}$ is the fraction of this dissipated energy which is injected into relativistic electrons. 
We assume that $\epsilon_\mathrm{e}$ is close to the equipartition value $\epsilon_\mathrm{e}=1/3$, as it is a necessary condition to have an efficient mechanism; ({\it iii}) $f_\mathrm{rad}$ is the fraction of the electron energy which is radiated. To explain the observed variability timescales in GRB lightcurves and to maintain a reasonable efficiency, 
the relativistic electrons must be in the fast cooling regime, which means that their radiative timescale is very short compared to the hydrodynamical timescales in the outflow. In this case, we have $f_\mathrm{rad}\simeq 1$.\\

The GRB LF in the internal shock model is therefore related to the physics of the relativistic ejection by the central engine.
We assume that the contrast $\kappa$ is distributed between $\kappa_\mathrm{min}$ and $\kappa_\mathrm{max}$ with a density of probability $\psi\left(\kappa\right)$ and that $\dot{E}$ is distributed between $\dot{E}_\mathrm{min}$ and $\dot{E}_\mathrm{max}$ with a density of probability $\phi\left(\dot{E}\right)$. The minimum and maximum radiated luminosities are therefore
\begin{equation}
L_\mathrm{min}= f\left(\kappa_\mathrm{min}\right) \dot{E}_\mathrm{min}
\end{equation}
 and 
\begin{equation}
L_\mathrm{max}= f\left(\kappa_\mathrm{max}\right) \dot{E}_\mathrm{max}\ .
\end{equation}
For $L_\mathrm{min}\le L \le L_\mathrm{max}$, the probability to have a 
radiated luminosity in the interval $\left[L;L+dL\right]$ is $p_0(L)dL$, 
where the intrinsic LF $p_0(L)$ is given by
\begin{equation}
p_0\left(L\right) = \int_{\max{\left(\kappa_\mathrm{min};
\stackrel{-1}{f}\left(\frac{L}{\dot{E}_\mathrm{max}}\right)\right)}}
^{\max{\left(\kappa_\mathrm{max};\stackrel{-1}{f}\left(\frac{L}
{\dot{E}_\mathrm{min}}\right)\right)}} \ \frac{\psi\left(\kappa\right)}{f\left(\kappa\right)}
\phi\left(\frac{L}{f\left(\kappa\right)}\right)d\kappa\ .
\label{eq:varphil}
\end{equation}

\subsection{The case of a power-law distribution of kinetic energy flux}
We assume that the injection rate of kinetic energy in the relativistic 
outflow follows a power-law distribution
\begin{equation}
\phi\left(\dot{E}\right) \simeq \frac{\delta-1}{\dot{E}_\mathrm{min}} \left(\frac{\dot{E}}{\dot{E}_\mathrm{min}}\right)^{\delta}\ ,
\end{equation}
with $\dot{E}_\mathrm{max} \gg \dot{E}_\mathrm{min}$. Then, the GRB LF given by equation~(\ref{eq:varphil}) becomes
\begin{eqnarray}
\lefteqn{p_0\left(L\right)  \simeq  \frac{\delta-1}{L_\mathrm{*}}\left(\frac{L}{L_\mathrm{*}}\right)^{-\delta}\times}\nonumber\\
& &  \int_{\max{\left(\kappa_\mathrm{min};\stackrel{-1}{f}\left(\frac{L}
{\dot{E}_\mathrm{max}}\right)\right)}}^{\min{\left(\kappa_\mathrm{max};
\stackrel{-1}{f}\left(\frac{L}{\dot{E}_\mathrm{min}}\right)\right)}} 
\ \psi(\kappa) \left(\frac{f\left(\kappa\right)}{f\left(\kappa_\mathrm{max}
\right)}\right)^{\delta-1}\ d\kappa,
\label{eq:varphilpl}
\end{eqnarray}
where the luminosity $L_\mathrm{*}$ is defined by
\begin{equation}
L_\mathrm{*} =  f\left(\kappa_\mathrm{max}\right) \dot{E}_\mathrm{min}\ .
\end{equation}

Let us now consider a first case where GRB central engines can produce all kinds of outflows, from highly variable to perfectly smooth. In this case, the minimum contrast is $\kappa_\mathrm{min}=1$, corresponding to a minimum efficiency $f\left(\kappa_\mathrm{min}\right)=0$, as no internal shocks can develop in an outflow with a constant Lorentz factor. The first limit in the integral in equation~(\ref{eq:varphilpl}) is then always given by $\stackrel{-1}{f}\left(L/\dot{E}_\mathrm{max}\right)$ and 
the LF is made of two branches :
\begin{itemize}
\item \textbf{High-luminosity branch.} 
For $L_\mathrm{*}\le L \le L_\mathrm{max}$, the second limit in the integral 
is $\kappa_\mathrm{max}$, which leads to
\begin{eqnarray}
p_0\left(L\right) & \simeq & \frac{\delta-1}{L_\mathrm{*}} \left(\frac{L}{L_\mathrm{*}}\right)^{-\delta}
f\left(\kappa_\mathrm{max}\right)\times\nonumber\\
& &
\int_{L/L_\mathrm{max}}^{1} dx\ x^{\delta-1}
\frac{ \psi\left(\stackrel{-1}{f}(x f\left(\kappa_\mathrm{max}\right))\right)}
{f'\left(\stackrel{-1}{f}(x  f\left(\kappa_\mathrm{max}\right))\right)}
\ .
\end{eqnarray}
For $L_\mathrm{*}\le L \ll L_\mathrm{max}$, $L/L_\mathrm{max}\to 0$ so that 
the integral is nearly constant. The high-luminosity branch of the LF is therefore very close to a power-law of slope $-\delta$, i.e. the slope of the intrinsic distribution of injected kinetic power.
\item \textbf{Low-luminosity branch.} For $L_\mathrm{min}=0 \le L \le L_\mathrm{*}$, the second limit in the integral is $\stackrel{-1}{f}\left(L/\dot{E}_\mathrm{min}\right)$, which leads to
\begin{eqnarray}
p_0\left(L\right) & \simeq & \frac{\delta-1}{L_\mathrm{*}} \left(\frac{L}{L_\mathrm{*}}\right)^{-\delta}
f\left(\kappa_\mathrm{max}\right)\times\nonumber\\
& &
\int_{L/L_\mathrm{max}}^{L/L_\mathrm{*}} dx\ x^{\delta-1}
\frac{ \psi\left(\stackrel{-1}{f}(x f\left(\kappa_\mathrm{max}\right))\right)}{f'\left(\stackrel{-1}{f}(x  f\left(\kappa_\mathrm{max}\right))\right)}
\ .
\end{eqnarray}
The LF is determined by the behaviour of the efficiency 
$f\left(\kappa\right)$ at very low contrast. We assume that 
$f\left(\kappa\right)\simeq f_{0}\left(\kappa-1\right)^{\alpha}$ for 
$\kappa\to 1$ and we define $L_{0}=f_{0}\dot{E}_\mathrm{min}$. Then, for $L\ll L_\mathrm{*}$, we have
\begin{eqnarray}
p_0\left(L\right) & \simeq & \frac{\delta-1}{L_{0}} \left(\frac{L}{L_{0}}\right)^{\frac{1}{\alpha}-1}
\frac{\psi\left(1\right)}{1+\alpha\left(\delta-1\right)}
\ .
\end{eqnarray}
We {therefore find that for $L\ll L_\mathrm{*}$ and
$\kappa_\mathrm{min}=1$, the LF is also a power-law}, with however a slope $1/\alpha-1$ which does not depend on the slope of the intrinsic distri\-bution of injected kinetic power.
\end{itemize}
{Then}, the predicted intrinsic GRB LF in the internal shock model is a broken power-law, of slope $1/\alpha-1$ at low-luminosity 
and $-\delta$ at high luminosity, with a break luminosity $L_\mathrm{*}=f\left(\kappa_\mathrm{max}\right)\dot{E}_\mathrm{min}$. 
The shape of $p_0(L)$ at the transition ($L\sim L_\mathrm{*}$) is related 
to the distribution of the contrast $\psi(\kappa)$.
{If $\kappa_\mathrm{min}\ne 1$ this result remains 
valid as long as very low contrasts 
can be achieved ($\kappa_\mathrm{min}\la 1.1$; see
figure~\ref{fig:philtheo}, left panel)}. Note that the analysis 
of the internal shock model parameter space made by \citet{barraud:05} 
shows that very low contrasts are necessary to produce soft GRBs such 
as X-ray rich GRBs (XRRs) and X-ray flashes (XRFs). However, if it happens 
that GRB central engines never produce smooth outflows
(i.e. if $\kappa_\mathrm{min}$ is not close to unity), the calculation made 
above remains valid for the high-luminosity branch, which is still a power-law 
of slope $-\delta$, but the low-luminosity branch is much reduced and no more 
a power-law.\\
{Finally, we have briefly considered the case where the
distribution of injected kinetic power $\phi(\dot{E})$ is not a
power-law. For example, for a log-normal distribution peaking at
$\dot{E}_\mathrm{*}$, we again obtain that the instrinsic LF follows
$\phi(\dot{E})$ at high luminosity and is a power-law of slope $-0.5$ at
low-luminosity, with a transition at $L_\mathrm{*}\sim f\left(\kappa_\mathrm{max}\right)\dot{E}_\mathrm{*}$.}\\

\subsection{A simple model for the internal shock efficiency}
To investigate more precisely the GRB LF, we need to know 
the form of the efficiency $f\left(\kappa\right)$. As shown in 
\citet{daigne:98}, it is a priori a complex function of the initial 
distribution of the Lorentz factor and the kinetic energy in the relativistic 
outflow. However, one can make a simple estimate by using the toy model 
developed in \citet{daigne:03,barraud:05} where we only consider direct 
collisions between two equal-mass relativistic shells. In this case, 
the efficiency is simply given by
\begin{equation}
f\left(\kappa\right) \simeq \epsilon_\mathrm{e}\times \frac{\left(\sqrt{\kappa}-1\right)^{2}}{\kappa+1}\ .
\end{equation}
For low contrast, it behaves as $f\left(\kappa\right)\simeq \epsilon_\mathrm{e}\left(\kappa-1\right)^{2}/8$, which corresponds to $f_{0}=\epsilon_\mathrm{e}/8$ and $\alpha=2$.
{In addition to this toy model which gives an explicit expression 
for the efficiency we have used our detailed internal shock code
\citep{daigne:98} and checked the quadratic dependence of $f(\kappa)$ in $(\kappa-1)$. 
Also notice that the result $\alpha=2$ will remain valid even if
$\epsilon_\mathrm{e}$
is not strictly constant, as long as it does not vary as some power of $(\kappa-1)$ at low
$\kappa$ ($\epsilon_e\propto (\kappa-1)$ leading for example to
$\alpha=3$ and
$p_0(L)\propto L^{-2/3}$ at low luminosity).}

\begin{figure*}
\centerline{\includegraphics[width=0.5\linewidth]{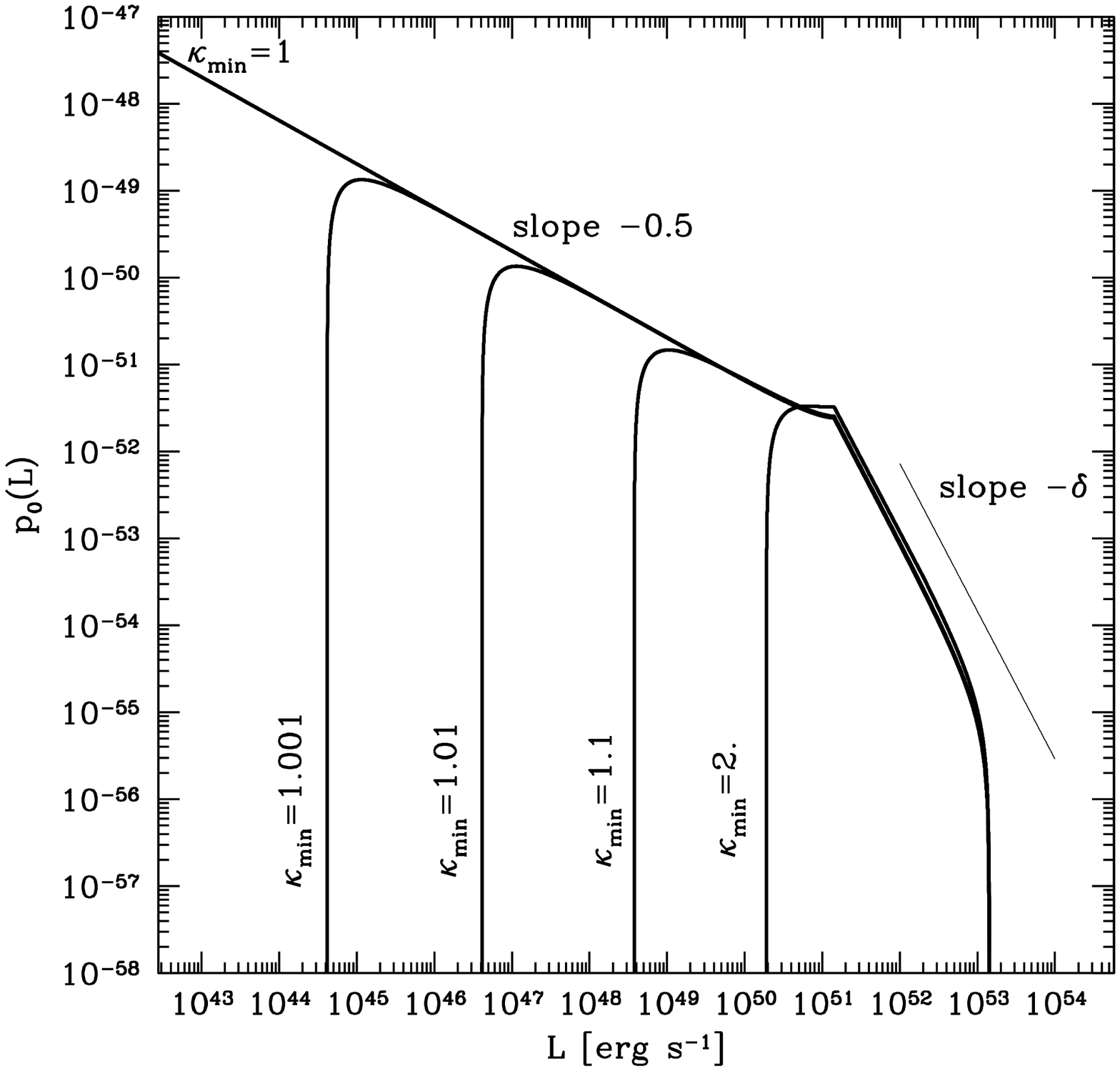}\includegraphics[width=0.5\linewidth]{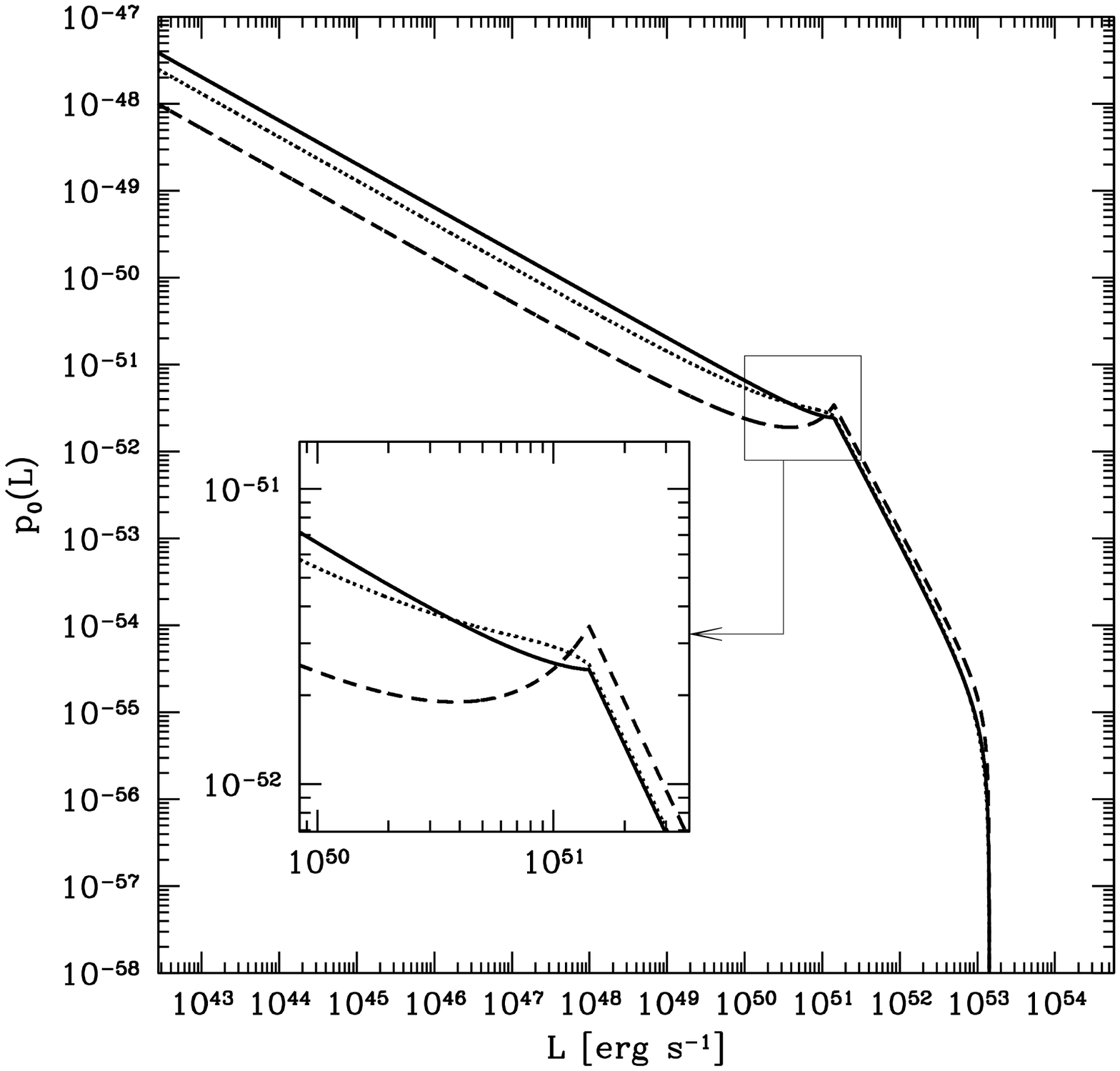}}
\caption{\textbf{The intrinsic GRB LF in the internal shock model.} 
The function $p_0(L)$ is plotted, assuming an intrinsic distribution of
 injected kinetic power $\dot{E}$ that is a power-law od slope
 $-\delta=-1.7$ between $\dot{E}_\mathrm{min}=10^{52}\ \mathrm{erg~s^{-1}}$ and $\dot{E}_\mathrm{max}=10^{54}\ \mathrm{erg~s^{-1}}$. 
This function $\phi(\dot{E})$ is plotted as a thin line in the left
 panel. \textit{Left: effect of the minimum constrast
 $\kappa_\mathrm{min}$.} We assume that $\log{\kappa}$ is uniformly
 distributed between $\log{\kappa_\mathrm{min}}$ and and
 $\log{\kappa_\mathrm{max}}$ with $\kappa_\mathrm{min}=1, 1.001, 1.01, 1.1$ and $2$, and $\kappa_\mathrm{max}=10$. \textit{Right panel: effect of the shape of the distribution of constrast.} 
We fix $\kappa_\mathrm{min}=1$ and $\kappa_\mathrm{max}=10$ and consider three different shapes for the distribution of constrast $\psi(\kappa)$~: (i) $\log{\kappa}$ uniformly distributed (solid line); (ii) $\psi(\kappa)\propto \exp{(-\kappa/4)}$ (dotted line); (iii) $\kappa$ uniformly distributed (dashed line). An inset shows the transition at $L_{*}=1.4\times 10^{51}\ \mathrm{erg~s^{-1}}$ in more details.
}
\label{fig:philtheo}
\end{figure*}

{Assuming a constant $\epsilon_e$,} we have plotted the resulting GRB 
LF in figure~\ref{fig:philtheo}. The intrinsic distribution 
of injected kinetic power is defined between $\dot{E}_\mathrm{min}=10^{52}\ \mathrm{erg~s^{-1}}$ 
and $\dot{E}_\mathrm{max}=10^{54}\ \mathrm{erg~s^{-1}}$ and has a slope
$-\delta=-1.7$.  
We have fixed the maximum value of the contrast to
$\kappa_\mathrm{max}=10$, so that $L_\mathrm{*}\simeq 1.4\times 10^{51}\
\mathrm{erg~s^{-1}}$. In the left panel, we have assumed that the logarithm of the contrast $\kappa$ is uniformly 
distributed between $\log{\kappa_\mathrm{min}}$ and 
$\log{\kappa_\mathrm{max}}$, with $\kappa_\mathrm{min}=1, 1.001, 1.01,
1.1$ or $2$. For all values 
of $\kappa_\mathrm{min}$, the high-luminosity branch is the same power-law 
of slope $-\delta=-1.7$. For $\kappa_\mathrm{min}=1$, the low-luminosity 
branch extends down to $L=0$ and is the expected power-law of 
slope $1/\alpha-1=-1/2$. This branch is still clearly visible 
for $\kappa_\mathrm{min}=1.001$, $1.01$ and $1.1$ but 
nearly disappears for $\kappa_\mathrm{min}=2$. For even higher values of $\kappa_\mathrm{min}$, only the high-luminosity power-law remains.\\
We have tested {in the right panel of figure~\ref{fig:philtheo}} that for other choices of the distribution of contrast 
$\psi(\kappa)$, the GRB LF is not affected (the two slopes 
remain unchanged) except for the shape of the transition at $L\sim L_\mathrm{*}$. 
A low-luminosity branch of slope $\sim -0.5$ in the intrinsic LF
is therefore a robust prediction 
of the internal shock model, as long as 
GRB central engines can produce 
smooth outflows (very low contrasts). 
{The low-luminosity branch will however manifest itself  
only if $L_*=f(\kappa_\mathrm{max}){\dot E}_{\mathrm{min}}$
is large enough; otherwise {the observationally accessible part
of} the LF of cosmological GRBs will behave as a
single power-law {(in figure~\ref{fig:philtheo},  $L_*\simeq
1.4\times 10^{51}\ \mathrm{erg~s^{-1}}$)}.}

\subsection{Apparent GRB luminosity function}
The GRB LF that has been derived from the internal
shock model is intrinsic. If GRB ejecta 
have a jet-like geometry with an opening angle $\Delta\theta$ which
is not correlated to the kinetic energy flux $\dot{E}$, the apparent
LF above $L_\mathrm{*}$ has the same shape as the
intrinsic one 
since the fraction of observed GRBs does not depend on $\dot E$.
At lower luminosities, two effects are in
competition : low-luminosity bursts can be due to a low internal shock
efficiency and/or a large viewing angle. Close to $L_\mathrm{*}$, the
first effect dominate and the slope is still very close to $-0.5$ as
predicted above. At very low luminosity, the
second effect takes over. It can be shown that the slope then becomes
close to $-7/6$. Observing this final slope seems difficult as it involves the
detection of very faint bursts. 
However viewing angle effects already modify the low-luminosity slope
below the break where it progressively departs from its 
intrinsic value $-0.5$ (see Fig.2).
\begin{figure*}
\centerline{\includegraphics[width=0.5\linewidth]{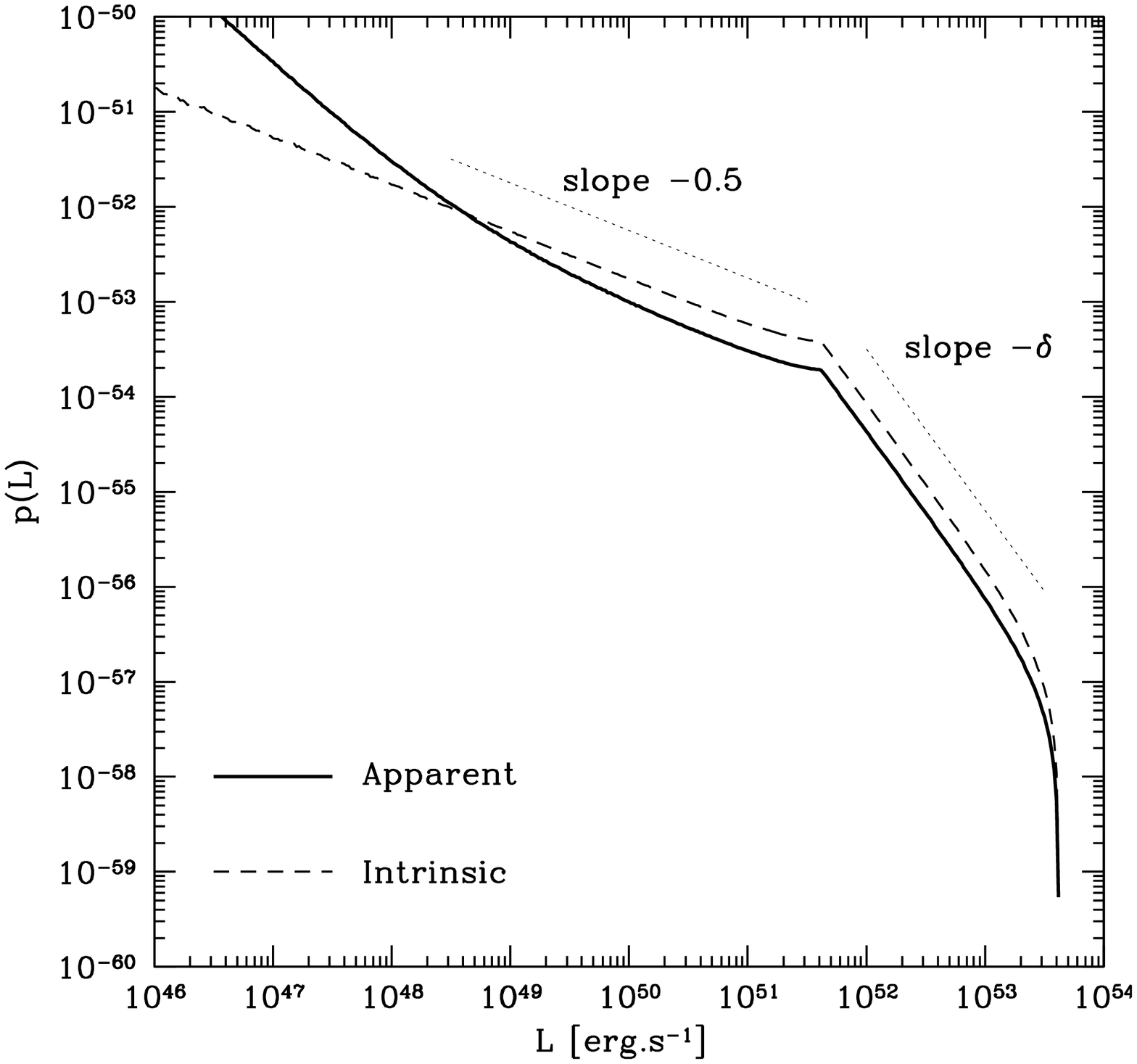}
\includegraphics[width=0.5\linewidth]{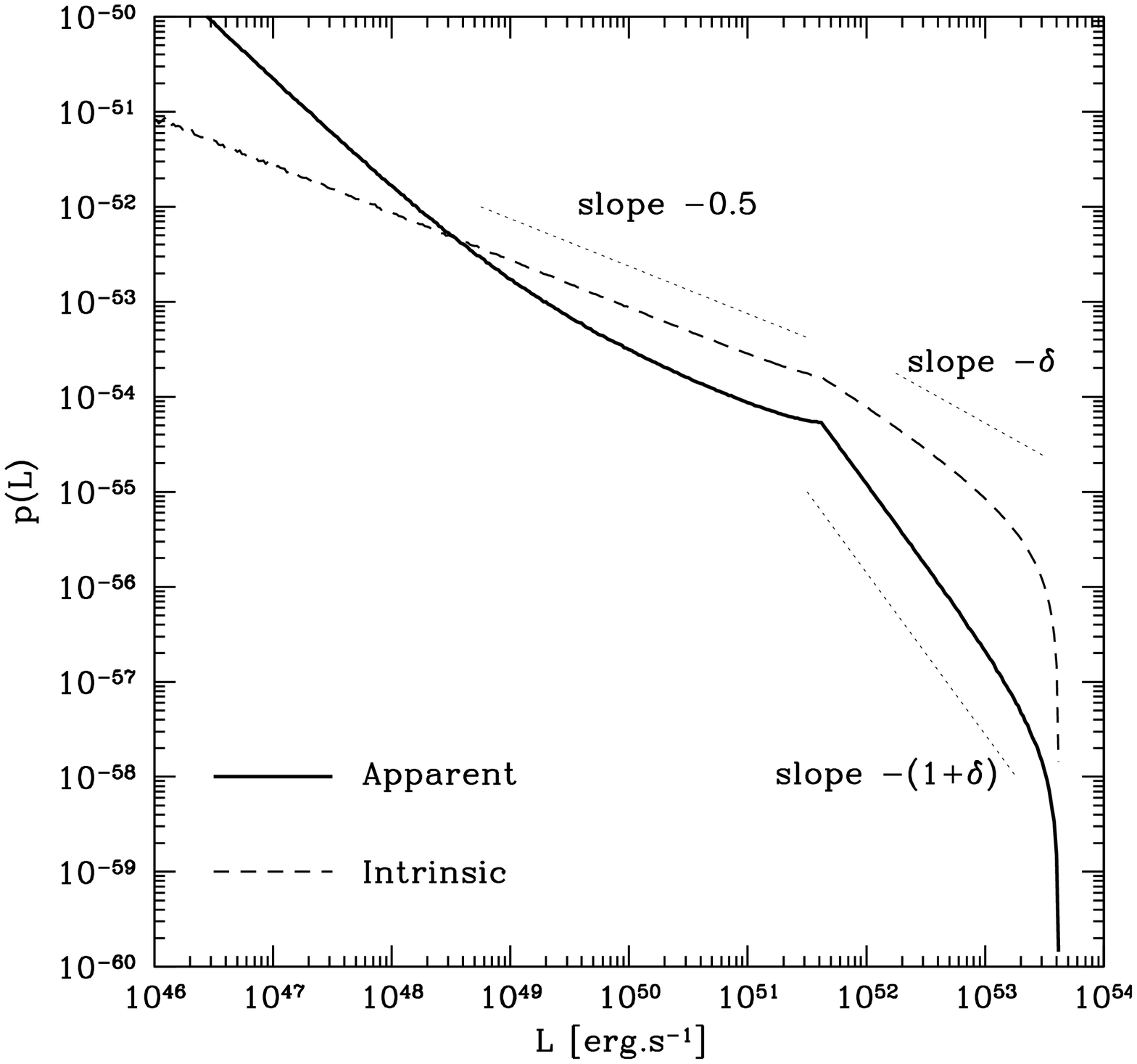}}
\caption{\textbf{The apparent GRB LF in the internal shock
 model.} The intrinsic LF $p_0(L)$ is plotted in dashed
 line for $\log(\kappa)$ uniformly distributed with
 $\kappa_\mathrm{min}=1$ and $\kappa_\mathrm{max}=10$, and for $\dot{E}$
 following a power-law distribution of slope $-\delta$ between
 $\dot{E}_\mathrm{min}=10^{52}\ \mathrm{erg~s^{-1}}$ and
 $\dot{E}_\mathrm{max}=10^{54}\ \mathrm{erg~s^{-1}}$. The corresponding
 apparent LF is plotted in solid line. \textit{Left:}
 the opening angle is distributed between $0$ and $\pi/2$ according to
 $p\left(\Delta\theta\right)=\sin{\Delta\theta}$ (uniformly distributed
 opening angle). We adopt  $\delta=-1.7$. \textit{Right:} the opening
 angle is correlated with the kinetic energy flux according to
 Eq.~(\ref{eq:correl}). We adopt $\delta=-0.7$.}
\label{fig:Lobs}
\end{figure*}

If it happens that the opening angle $\Delta\theta$ is correlated with
$\dot{E}$, the apparent $p(L)$ will be  also different from the intrinsic one
at high luminosity.
A possible correlation could be 
\begin{equation}
\dot{E}\left(1-\cos{\Delta\theta}\right) = \mathrm{cst} = \dot{E}_\mathrm{min}\ .
\label{eq:correl}
\end{equation}
meaning that the true kinetic energy rate is the same for
all bursts. Such an assumption is motivated by the
evidence that there might be a standard energy reservoir in GRBs  
\citep{frail:01}. 
In this case, the high-luminosity
branch (above $L_\mathrm{*}$) of $p(L)$ has a slope $-\left(1+\delta\right)$,
where $-\delta$ is the slope of the intrinsic LF. This
is illustrated in figure~\ref{fig:Lobs} (right panel).

\section{Constraining the GRB luminosity function}
\subsection{Monte Carlo simulations}
We use Monte Carlo simulations to constrain the GRB LF from observations. The method is described in
details in \citet{daigne:06}. We recall here the main lines~:
\begin{itemize}
\item \textbf{Properties of the long GRB population.}
We characterize this population by the intrinsic distribution of several physical properties :
(1) the comoving rate. We define $\mathcal{R}_\mathrm{GRB}(z)$ ($\mathrm{Mpc^{-3}~yr^{-1}}$) as the GRB comoving rate at redshift $z$.  We assume that the GRB comoving rate is 
proportional to the star formation rate (SFR). Following \citet{porciani:01}, 
this can be written as $\mathcal{R}_\mathrm{GRB}=k\times \mathcal{R}_\mathrm{SN}$, 
where $\mathcal{R}_\mathrm{SN}$ is the comoving rate of collapses of massive 
stars above 8 $M_\mathrm{\odot}$. We consider three possible scenarios 
(see Fig.1 in \citet{daigne:06} {and reference therein}), that all fit the observed SFR up to 
$z\sim 2-3$ : SFR$_{1}$ where the SFR decreases for $z\ga 2$, SFR$_{2}$ where 
it is constant for $z\ga 2$ and SFR$_{3}$ where it increases for $z \ga 2$. 
In this last case, we have to assume a maximum redshift for star formation. 
We adopt $z_\mathrm{max}=20$; 
(2) the LF. In this paper, {for simplicity,} we do not discuss evolutionary 
effects. 
Therefore, the probability density $p(L)$ of the isotropic equivalent 
luminosity $L$ does not depend on $z$. In \citet{daigne:06} we only considered 
the case of a power-law distribution, with $p(L)\propto L^{-\delta}$ for $L_\mathrm{min}\le L \le L_\mathrm{max}$. Here, we test more complicated LFs, i.e. broken power-laws defined by
\begin{equation}
p(L) \propto \left\lbrace\begin{array}{cl}
L^{-\delta_{1}} & \mathrm{for}\ L_\mathrm{min}\le L\le L_\mathrm{b}\ ,\\
L^{-\delta_{2}} & \mathrm{for}\ L_\mathrm{b}  \le L\le L_\mathrm{max}\ ;\\
\end{array}\right.
\end{equation}
 (3) the distributions of intrinsic spectral parameters. We assume that the GRB photon spectrum is given by a broken power-law with a break at energy $E_\mathrm{p}$ \citep{band:93} and a slope $-\alpha$ (resp. $-\beta$) at low (resp. high) energy. We checked that the use of the more realistic spectrum shape proposed by \citet{band:93} does not affect our conclusions.
The slopes $\alpha$ and $\beta$ are given the distributions observed 
in a sample of BATSE bright long GRBs \citep{preece:00}. 
For the peak energy $E_\mathrm{p}$ we consider two possible cases, either a log-normal distribution, with a mean value $E_\mathrm{p,0}$ and a dispersion $\sigma=0.3\ \mathrm{dex}$ (hereafter ``log-normal $E_\mathrm{p}$ distribution'') or an intrinsic correlation between the spectral properties and the luminosity (hereafter ``Amati-like relation''), as found by \citet{amati:02,amati:06}. We assume in this case that
\begin{equation}
E_\mathrm{p}=380\ \mathrm{keV}\ \left(\frac{L}{1.6\times 10^{52}\ \mathrm{erg~s^{-1}}}\right)^{0.43}\ ,
\label{eq:amati}
\end{equation}
with a normal dispersion $\sigma=0.2\ \mathrm{dex}$ in agreement with observations (\citet{yonetoku:04,ghirlanda:05}, see however \citet{nakar:05,band:05} who tested this relation against BATSE data and concluded that selection effects were dominant). 
\item \textbf{Criteria of detection by several instruments.} With
the assumptions listed above, a GRB in the simulation is characterized by a 
redshift $z$, a peak luminosity $L$, and a spectrum defined by $E_\mathrm{p}$, 
$\alpha$ and $\beta$. It is therefore possible to compute the observed peak 
flux in any spectral band. Using the known sensitivity of several instruments 
(see \citet{daigne:06} for the detailed thresholds we use), we can determine 
if a given burst is detected by the following experiments : ({\it i}) BATSE, 
for which we define two samples in our synthetic bursts (all BATSE bursts 
and bright BATSE bursts) ; ({\it ii}) HETE2, for which we test the detection 
either by the gamma-ray (FREGATE) or the X-ray (WXM) instrument; 
and ({\it iii}) SWIFT, for which we test the detection by the gamma-ray instrument 
only (BAT) and we also define two samples (all SWIFT bursts and 
bright SWIFT bursts). It is then possible to compute simulated observed 
distribution of various quantities to compare them with real data.
\item \textbf{Observational constrains.} We use three different kind 
of observations: (1) the $\log{N}-\log{P}$ diagram of BATSE bursts 
\citep{stern:00,stern:01}; (2) the observed peak energy distribution of long 
bright GRBs \citep{preece:00}; and (3) the observed fraction of soft GRBs 
(X-ray Flashes and X-ray rich GRBs) in the sample of GRBs detected 
by HETE2 \citep{sakamoto:05}. The $\log{N}-\log{P}$ diagram is broadly used 
for such kind of studies but we have shown in \citet{daigne:06} 
that (2) and (3) are good complements to better constrain the parameters 
of the GRB population.
\end{itemize}
Depending on the assumptions on the spectral properties, we have 4 or 5
free parameters for a single power-law LF ($k$, $L_\mathrm{min}$,
$L_\mathrm{max}$, $\delta$ and $E_\mathrm{p,0}$ in the case of a
log-normal distribution). The observational constraints correspond to 41
data points {(see figure~\ref{fig:fit} : 30 data points in the $\log{N}-\log{P}$ diagram
published by \citet{stern:01}; 10 points for the $E_\mathrm{p}$
distribution published by \citet{preece:00} which has been rebinned in 10
logarithmic bins of size 0.2 dex between 15.8 keV and 1.58 MeV; 1 point
for the fraction of soft GRBs)}. We have therefore 37 or 36 degrees of freedom.
The numerical procedure is the following : for a set of parameters, 
we generate randomly a population of $10^{5}$ GRBs using the distribution defined above, we then compute the simulated distributions of observed peak flux, peak energy, etc. and we compare them to real data, by computing a $\chi^{2}$. We do that for a large number of sets of parameters, randomly chosen to explore a large space. We always find a clear minimum $\chi^{2}_\mathrm{min}$ of $\chi^{2}$ and we define as ``best models'' all models with $\chi^{2}_\mathrm{min}\le \chi^{2}\le \chi^{2}_\mathrm{min}+\Delta\chi^{2}$, where $\Delta\chi^{2}$ defines the $1\sigma$ level and is computed from the number of degrees of freedom.\\

The main results obtained in \citet{daigne:06} are (1) that SFR$_{3}$ is 
strongly favored by the observed redshift distribution of SWIFT bursts. 
But a SFR rising at large $z$ appears unlikely as it would overproduce
metals at early times. This is therefore an indirect indication
in favor of 
a GRB rate that does not directly 
follow the SFR, for instance due to an evolution with redshift of the 
efficiency of massive stars to produce GRBs\footnote{{In \citet{daigne:06}, we have tested whether an evolution of the LF could reconcile the
      SFR$_{1}$ or SFR$_{2}$ scenario with Swift data. We found that this is very
      unlikely as the evolution should be strong ($L\propto (1+z)^{\nu}$ with $\nu > 2$). We therefore conclude
that the evolution of the GRB rate (i.e the evolution of the stellar
      efficiency to produce GRBs) is dominant compared to a possible
      evolution of the LF.}}; (2) with this SFR$_{3}$, 
both the ``Amati-like relation'' and the ``log-normal $E_\mathrm{p}$ 
distribution'' give good fits to the observational constraints listed above.
Best model parameters 
of the LF do not vary too much from one scenario to another : 
the slope is well determined, $\delta\sim 1.5-1.7$, but the minimum and 
maximum luminosities are not so well constrained, with 
$L_\mathrm{min}\sim 0.8-3\times 10^{50}\ \mathrm{erg~s^{-1}}$ 
and $L_\mathrm{max}\sim 3-5\times 10^{53}\ \mathrm{erg~s^{-1}}$. 
It is interesting to note that with a different methodology, several groups 
have confirmed our conclusions on the GRB comoving rate \citep{le:07,guetta:07,kistler:07}.\\

In this paper, we present the results of additional simulations that we have 
carried out to test if the GRB
LF can be a broken power-law. With two supplementary parameters (the break luminosity $L_\mathrm{b}$ and
the second slope), we have now 6 (resp. 7) free parameters in the Amati-like relation case (resp. the case of a log-normal peak energy distribution). 
It is difficult to constrain accurately so many parameters with Monte Carlo simulations. Therefore, we have chosen to keep the maximum luminosity constant in all our simulations. We adopt $L_\mathrm{max}=10^{53.5}\ \mathrm{erg~s^{-1}}$, according to our previous study \citep{daigne:06}. We also keep $L_\mathrm{min}$ constant, and equal to a low value corresponding to weak GRBs that cannot be detected at cosmological distance. We usually adopt $L_\mathrm{min}=10^{45}\ \mathrm{erg~s^{-1}}$ but we have also tested other values (see next section). Keeping these two luminosities constant in our Monte Carlo simulation, 
we have the same number of degrees of freedom than in the model with a simple power-law LF.


\begin{table*}
 \centering
\begin{minipage}{172mm}
\caption{Best models: parameters.\label{tab:bestmodels}}
\begin{tabular}{ccccccccc}
  \hline
  {SFR} & $\log{L_\mathrm{min}}$  & $\log{L_\mathrm{b}}$    & $\log{L_\mathrm{max}}$  & $\delta_1$ & $\delta_2$  & $\log{E_\mathrm{p,0}}$ & $\log{k}$ & $\log{\rho_{0}}$ \\
               & ($\mathrm{erg~s^{-1}}$) & ($\mathrm{erg~s^{-1}}$) & ($\mathrm{erg~s^{-1}}$) &            &             & (keV)                  &           & ($\mathrm{GRB~Gpc^{-3}~yr^{-1}}$)\\
  \hline
  \multicolumn{8}{c}{\textbf{Amati-like relation $E_\mathrm{p}\propto L^{0.43}$}}\\
3\footnote{Preferred model in the case of a single power-law LF \citep{daigne:06}.} 
        & $50.3\pm 0.7$ &               & $53.5\pm 0.4$ & $1.54\pm 0.18$ &                & & $-6.0\pm 0.2$ & $-0.8\pm 0.2$ \\
1       & $45$          & $50.4\pm 0.5$ & $53.5$        & $0.65\pm 0.22$ & $1.71\pm 0.07$ & & $-5.2\pm 0.3$ & $~~0.0\pm 0.3$ \\
2       & $45$          & $50.5\pm 0.4$ & $53.5$        & $0.62\pm 0.20$ & $1.71\pm 0.09$ & & $-5.3\pm 0.2$ & $-0.1\pm 0.2$ \\
3       & $45$          & $51.2\pm 0.6$ & $53.5$        & $0.60\pm 0.22$ & $1.71\pm 0.22$ & & $-5.9\pm 0.2$ & $-0.7\pm 0.2$ \\
3       & $46$          & $51.1\pm 0.7$ & $53.5$        & $0.64\pm 0.20$ & $1.70\pm 0.29$ & & $-5.9\pm 0.2$ & $-0.7\pm 0.2$ \\
3       & $47$          & $51.2\pm 0.6$ & $53.5$        & $0.69\pm 0.24$ & $1.70\pm 0.30$ & & $-5.8\pm 0.2$ & $-0.6\pm 0.2$ \\
3       & $48$          & $51.3\pm 0.5$ & $53.5$        & $0.74\pm 0.27$ & $1.75\pm 0.38$ & & $-5.8\pm 0.2$ & $-0.6\pm 0.2$ \\
3       & $49$          & $51.6\pm 1.1$ & $53.5$        & $1.02\pm 0.45$ & $1.98\pm 0.72$ & & $-5.7\pm 0.2$ & $-0.5\pm 0.2$ \\
3       & $50$          & $52.0\pm 1.2$ & $53.5$        & $1.47\pm 0.65$ & $2.05\pm 0.86$ & & $-5.9\pm 0.1$ & $-0.7\pm 0.1$ \\
  \hline
  \multicolumn{8}{c}{\textbf{log-normal peak energy distribution}}\\
3$^{\displaystyle a}$
        & $50.5\pm 1.3$ &               & $53.7\pm 0.9$ & $1.52\pm 0.48$ &                & $2.79\pm 0.08$ & $-6.2\pm 0.2$ & $-1.0\pm 0.2$ \\
1       & $45$          & $51.2\pm 0.4$ & $53.5$        & $0.67\pm 0.23$ & $1.80\pm 0.16$ & $2.80\pm 0.11$ & $-5.5\pm 0.2$ & $-0.3\pm 0.2$ \\
2       & $45$          & $51.5\pm 0.5$ & $53.5$        & $0.71\pm 0.21$ & $2.04\pm 0.57$ & $2.75\pm 0.09$ & $-5.6\pm 0.3$ & $-0.5\pm 0.3$ \\
3       & $45$          & $52.1\pm 0.5$ & $53.5$        & $0.66\pm 0.20$ & $2.37\pm 0.71$ & $2.80\pm 0.08$ & $-6.1\pm 0.2$ & $-0.9\pm 0.2$ \\
 \hline
\end{tabular}
\end{minipage}
\end{table*}
\begin{figure*}[!p]
\centerline{\includegraphics[width=0.5\linewidth]{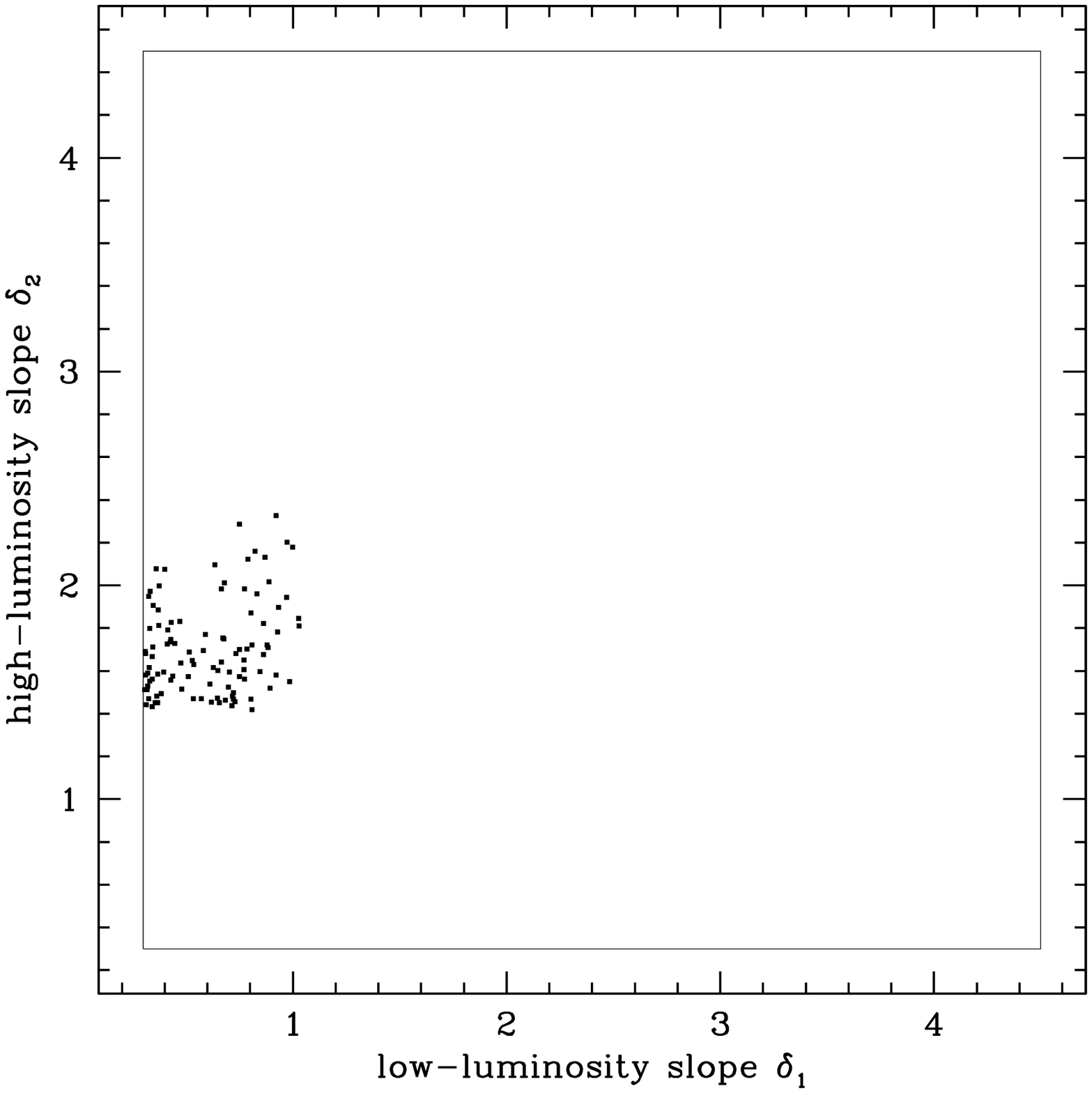}
\includegraphics[width=0.5\linewidth]{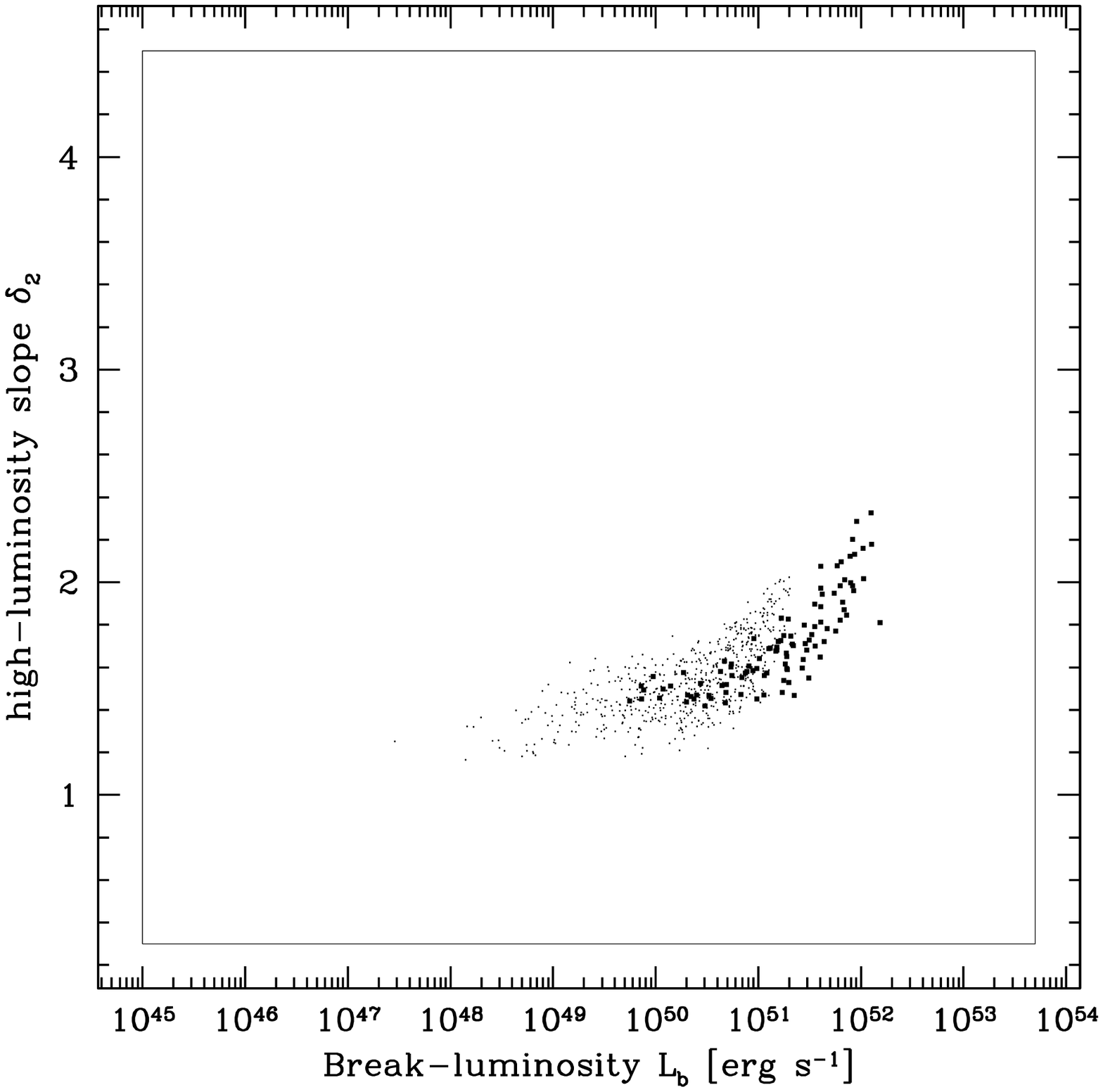}}
 \caption{\textbf{Parameter space (SFR$_{3}$, Amati-like relation) :} we
 plot the location of the best models ($1\ \sigma$ level) for a broken
 power-law LF with fixed minimum and maximum luminosities
 $L_\mathrm{min}=10^{45}\ \mathrm{erg~s^{-1}}$ and
 $L_\mathrm{max}=10^{53.5}\ \mathrm{erg~s^{-1}}$ (big dots). The range
 of parameters explored in the Monte Carlo is indicated with a box. In the right panel, the best models for a single power-law LF \citep{daigne:06} are also plotted with small dots (in this case the x-axis stands for the minimum luminosity and the y-axis for the slope).}
\label{fig:parameterspace}
\end{figure*}

\begin{figure*}[!p]
\centerline{\includegraphics[width=0.5\linewidth]{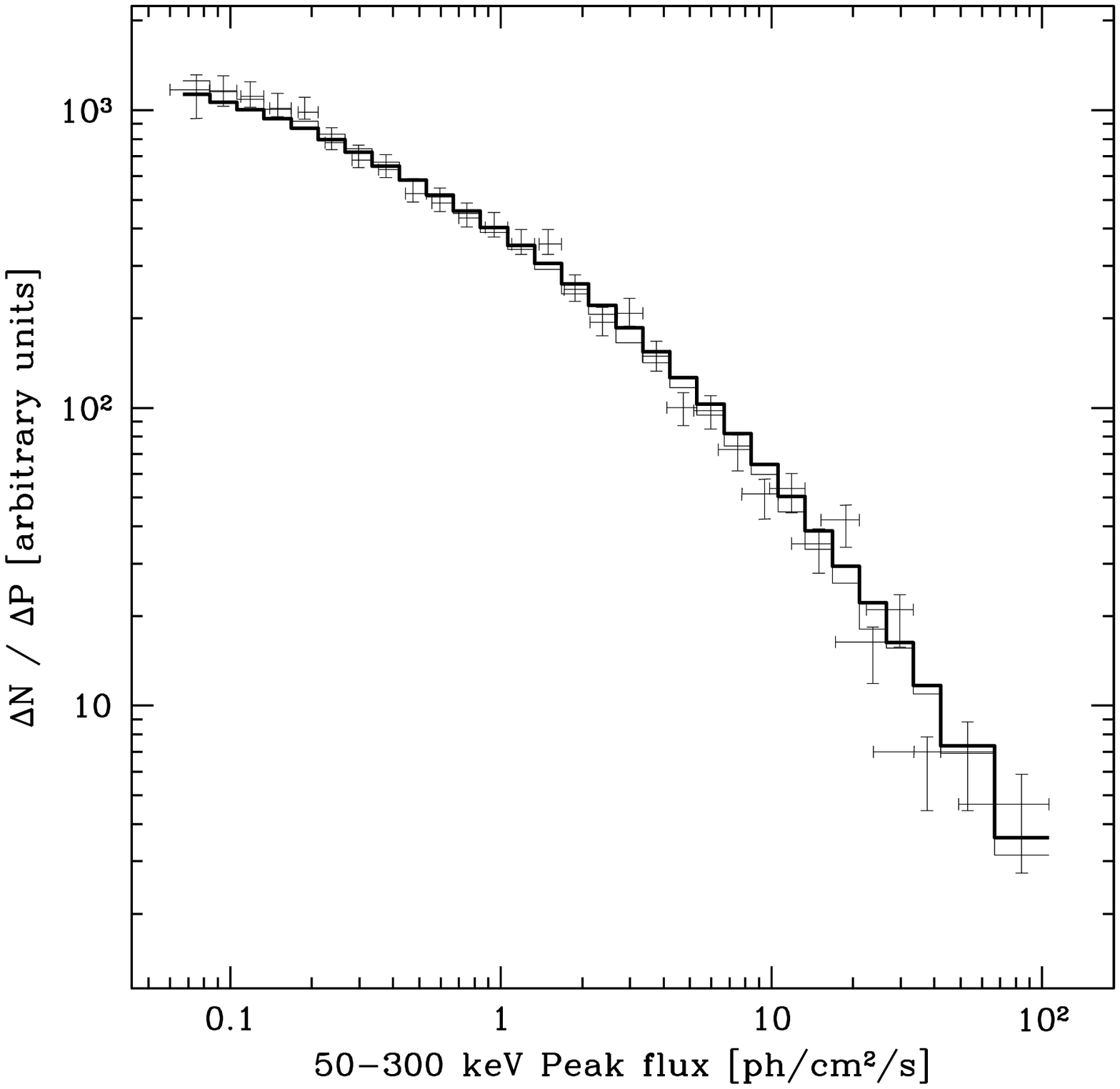}
\includegraphics[width=0.5\linewidth]{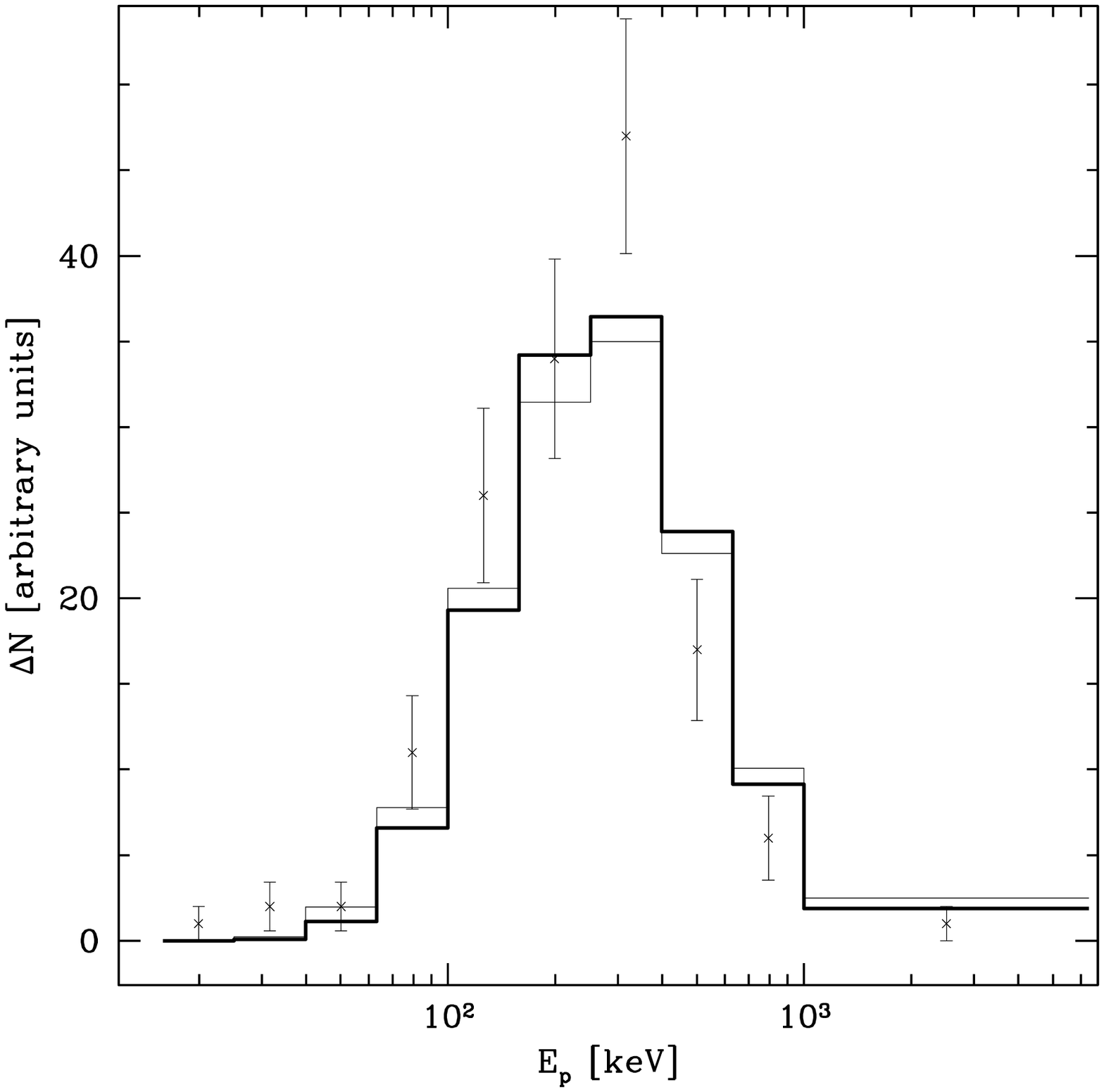}}
 \caption{\textbf{Best model (SFR$_{3}$, Amati-like relation).} \textit{Left:} the simulated $\log{N}-\log{P}$ diagram of BATSE is plotted as well as BATSE data \citep{stern:02a}; \textit{right:} the simulated peak energy distribution of bright BATSE bursts is plotted as well as the observed distribution \citep{preece:00}. In both panels, the best model for a broken power-law LF with fixed minimum and maximum luminosities  $L_\mathrm{min}=10^{45}\ \mathrm{erg~s^{-1}}$ and  $L_\mathrm{max}=10^{53.5}\ \mathrm{erg~s^{-1}}$ is plotted in thick line. For comparison the best model for a single power-law LF obtained in \citet{daigne:06} is plotted in thin line.}
\label{fig:fit}
\end{figure*}

\subsection{Results}
In our whole new set of simulations, we always find a clear minimum of $\chi^{2}$. The parameters of the best model, as well as $1\ \sigma$ error bars are listed in table~\ref{tab:bestmodels}. 
As can be seen, we focus on the scenario where the comoving GRB rate 
follows SFR$_{3}$ and the peak energy is given by the Amati-like relation.
For comparison, we also give the parameters of two reference models with a single power-law LF \citep{daigne:06}.
 Figure~\ref{fig:parameterspace} illustrates, in the case SFR$_{3}$+Amati-like relation, the position 
of the best models in the parameter space of the LF. 
As can be seen, the low-luminosity slope is strongly constrained to be small, 
$\delta_{1}\la 1$, with a mean value $\delta_{1}\sim 0.6$, while 
the high-luminosity slope is larger, $\delta_{2}\ga 1.4$, with a mean value $\delta_{2}\sim 1.7$. 
The break luminosity (right panel) 
is not so well constrained with best models having $L_\mathrm{b}\simeq 4\times 10^{50}$--$6\times 10^{51}\ \mathrm{erg~s^{-1}}$. 
Figure~\ref{fig:fit} compares the fit of the data points ($\log{N}-\log{P}$ 
diagram and $E_\mathrm{p}$ distribution) with the best model obtained either 
with a power-law or a broken power-law LF. Both models are in good agreement 
with data, without a preference for one or the other. This is also indicated 
by the value of the reduced minimum $\chi^{2}$ in both cases : 1.4 (power-law) 
and 1.3 (broken power-law) for 37 degrees of freedom.

\begin{figure}
\centerline{\includegraphics[width=\linewidth]{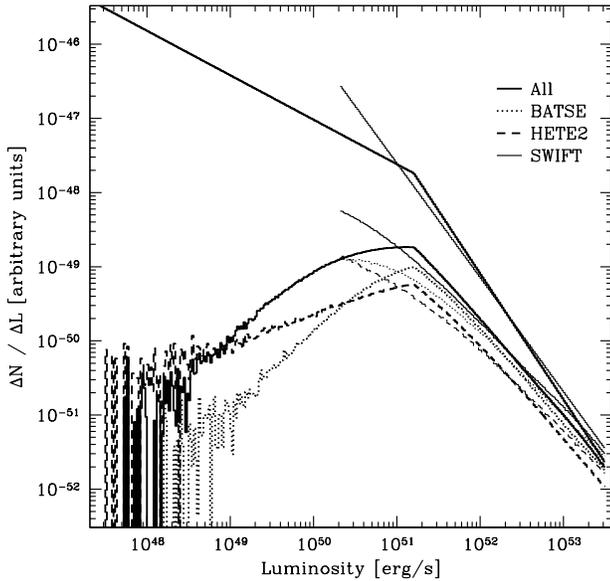}}
 \caption{\textbf{Luminosity function in the scenario 
SFR$_{3}$ + Amati-like relation.} The apparent LF, 
as well as the luminosity distribution of bursts detected by BATSE,
 HETE2 and SWIFT are plotted in thick line for the best model using a
 broken power-law with fixed minimum and maximum luminosities
 $L_\mathrm{min}=10^{45}\ \mathrm{erg~s^{-1}}$ and
 $L_\mathrm{max}=10^{53.5}\ \mathrm{erg~s^{-1}}$. All other parameters
 can be found in table~\ref{tab:bestmodels}. For comparison the best
 model for a single power-law LF obtained in \citet{daigne:06} is
 plotted in thin line. 
\textbf{Despite the fact that this figure was obtained 
with a special run simulating $10^{9}$ GRBs with the best model parameters, 
the curves are very noisy at low luminosity, as these events are very rare.}}
\label{fig:luminosity}
\end{figure}

Figure~\ref{fig:luminosity} shows -- for the best model -- the  
LF as well as the luminosity  distribution of bursts 
detected by BATSE, HETE2 and SWIFT.  It appears that the 
high-luminosity branch above $L_\mathrm{b}$ is extremely close 
to the best model single power-law LF (thin line). 
{Below a few $10^{49}\ \mathrm{erg~s^{-1}}$ (corresponding to the
lowest values of the constrast, $\kappa\la 1.2$), the fraction of detected GRBs is 
extremely low (less than $10^{-4}$ of the total). The two models (power-law vs broken power-law) 
 differ in the $10^{49}$--$10^{51}\ \mathrm{erg~s^{-1}}$ range, where
the fraction of detected bursts is still small {(less than 20\%
 of the total)} with therefore little
effect on the observable quantities.} \\

These results indicate that present data are not sufficient to distinguish 
between a power-law and a broken power-law LF. Both models 
can provide equally good fits to the observations. It is however interesting 
that a broken power-law remains allowed, as there are good
theoretical arguments in favor of such a shape (see Sect.2). Table~\ref{tab:bestmodels} 
show that the properties of the broken power-law LF remain very stable
as long as $L_\mathrm{min}$ 
is kept to a low value ($L_\mathrm{min}\la 10^{48}\ \mathrm{erg~s^{-1}}$): 
the position of the break and especially the values of the two slopes are not 
changing much, even for different GRB rates (SFR$_{1}$ and SFR$_{2}$ have 
also been tested). It seems however that the broken 
power-law LF is more sensitive to the assumptions on
the spectral parameters. In the case where the spectral properties are not 
correlated to the luminosity (log-normal peak energy distribution), 
the low-luminosity slope is not too different from the ``Amati-like relation'' case 
($\delta_{1}\simeq 0.7$ instead of $0.6$), but the break luminosity is 
larger ($L_\mathrm{b}\simeq 10^{51-52}$ instead of $10^{50-51}\ \mathrm{erg~s^{-1}}$) and the high-luminosity branch is steeper ($\delta_{2}\simeq 1.8-2.4$ instead of $1.7$).\\ 

These results can be partially compared to other studies. Based on an analysis of the BATSE $\log{N}-\log{P}$ diagram, \citet{stern:02b} have tested several shapes of GRB LFs, including a power-law and a broken power-law. Their assumptions concerning GRB spectra are different from those chosen in the present paper but are very close to our ``log-normal peak energy distribution'' scenario. 
For a GRB rate similar to our SFR$_{3}$, they find $\delta_{1}\simeq 1.3-1.6$, 
$\delta_{2}\ga 3$ and $L_\mathrm{b}\simeq 10^{51}-6\times 10^{52}\ \mathrm{erg~s^{-1}}$. 
Whereas we are in reasonable agreement for the high luminosity slope, there 
is a large discrepancy for the low-luminosity branch, which is much steeper 
in their study. To understand the origin of this discrepancy, we made an 
additional simulation where we force $\delta_{1}=1.3$ and let $\delta_{2}$ 
and $L_\mathrm{b}$ free. We find that a good fit to the $\log{N}-\log{P}$ 
diagram can be found but that the peak energy distribution is not reproduced, which stresses the importance of this additional constrain in our study. \citet{firmani:04} have presented a set of Monte Carlo simulations with assumptions on the intrinsic GRB properties that are very similar to ours but different observational constraints, as they fit the distribution of BATSE pseudo-redshifts obtained from the luminosity--variability correlation \citep{fenimore:00}. For the case with an intrinsic correlation between the luminosity and the spectral properties (Amati-like case) they find a break at $L_\mathrm{b}\simeq 3\times 10^{52}\ \mathrm{erg~s^{-1}}$ with the low and high-luminosity slopes equal to $\delta_{1}\simeq 0.8$ and $\delta_{2}\simeq 2.1$. Taking into account the differences 
in the two approaches, the agreement between this study and our result
is satisfactory, especially for the slopes. More recently, \citet{guetta:05,guetta:07} have also studied the GRB rate and LF using the recent results from SWIFT. Their analysis is based on the use of the $\log{N}-\log{P}$ diagram only and they assume a very simplified GRB spectral shape, that is a power-law spectrum of photon index $-1.6$. 
For SFR$_3$, they find a 
break luminosity $L_\mathrm{b}\simeq 4\times 10^{51}\ \mathrm{erg.s^{-1}}$ 
and low and high luminosity slopes $\delta_{1}\simeq 0.1$ and $\delta_{2}\simeq 2$, 
with large error bars. There is therefore a discrepancy for the low-luminosity 
slope which is flatter in their case. We believe however that the use of 
a more realistic spectral shape together with a constraint on the peak energy 
distribution leads in our case to a better determined LF at low luminosity.\\    

The two important results from this new set of Monte Carlo simulations 
are that (1) a broken power-law is compatible with present data but 
is not preferred compared to a single power-law (equally good fits of the
observations); (2) if the LF is indeed a broken power-law, 
the low-luminosity slope is constrained to be $\delta_{1}\simeq 0.6\pm 0.2$ 
(Amati-like relation) or $\delta_{1}\simeq 0.7\pm 0.2$ (log-normal peak energy 
distribution), i.e. compatible with the prediction of the internal shock model 
derived in Sect.2.\\

\subsection{The rate of underluminous GRBs}
A by-product of this study is an estimate of the local GRB rate, which
is given in table~\ref{tab:bestmodels} and is typically{, in the
SFR$_{3}$ scenario,} $0.1-0.3\ \mathrm{GRB~Gpc^{-3}~yr^{-1}}$ in the Amati-like case, and $0.08-0.2\ \mathrm{GRB~Gpc^{-3}~yr^{-1}}$ for a log-normal peak energy distribution.  This is in good agreement with the results of \citet{guetta:07}. Despite the fact that our broken power-law LF can in principle extend to very low-luminosity, such a low local rate corresponds to less that $10^{-3}\ \mathrm{GRB~yr^{-1}}$ within 100 Mpc, which cannot explain 
the observation of GRB 980425 at z=0.008 or GRB 060618 at z=0.03. 
As shown in \citet{daigne:07}, such underluminous bursts are well explained 
in the framework of the internal shock model by mildly relativistic / mildly 
energetic outflows. This would then indicate that the collapsing stars capable 
to generate such outflows, less extreme than those required to produce 
standard GRBs, are very numerous and should then produce an additional component 
in the LF, dominant at very low luminosity. 
This new branch cannot be the simple continuation of the LF of standard GRBs 
derived in this paper. 
Recently \citet{liang:06} have studied such a two component LF 
and found that the local rate corresponding to the low-luminosity component 
has to be several orders of magnitude above that of standard GRBs 
but can still represent only a fraction of all type Ib/c supernovae.

\section{Conclusion}
We have demonstrated that in the framework of the internal shock model, a 
two branch LF is naturally expected, with a predicted 
low-luminosity branch which is a power-law of slope close to $ -0.5$. 
This result 
is robust as long as the central engine responsible for GRBs is capable 
to produce a broad diversity of outflows, from highly variable to very smooth.\\

Using a set of Monte Carlo simulations, we have then shown that current 
observations ($\log{N}-\log{P}$ diagram, peak energy distribution, fraction 
of XRRs and XRFs) are compatible with a broken power-law LF
but still do not exclude  a single power-law distribution. 
The low-luminosity slope of the broken power-law is strongly constrained to be 
$\delta_{1}\simeq 0.4-0.9$, compatible with the prediction of the internal shock model.\\

These results are encouraging but only preliminary. A better 
determination of the GRB LF would provide 
an interesting 
test of the internal shock model when the low luminosity branch 
becomes more easily accessible. This will however require the difficult task of 
detecting many bursts at the
threshold of current instruments and measuring their
redshift and spectral properties.  \\

\section*{Acknowledgments}
HZ is supported by a BAF (Bourse Alg\'ero-Fran{\c c}aise) 
No. 40/ENS./FR/2006/2007.

\bibliographystyle{mn2e}
\bibliography{break}

\begin{thebibliography}{}

\bibitem[\protect\citeauthoryear{{Amati}}{{Amati}}{2006}]{amati:06}
{Amati} L.,  2006, \mnras, 372, 233

\bibitem[\protect\citeauthoryear{{Amati}, {Frontera}, {Tavani}, {in't Zand},
  {Antonelli}, {Costa}, {Feroci}, {Guidorzi}, {Heise}, {Masetti}, {Montanari},
  {Nicastro}, {Palazzi}, {Pian}, {Piro} \& {Soffitta}}{{Amati}
  et~al.}{2002}]{amati:02}
{Amati} L.,  {Frontera} F.,  {Tavani} M.,  {in't Zand} J.~J.~M.,  {Antonelli}
  A.,  {Costa} E.,  {Feroci} M.,  {Guidorzi} C.,  {Heise} J.,  {Masetti} N.,
  {Montanari} E.,  {Nicastro} L.,  {Palazzi} E.,  {Pian} E.,  {Piro} L.,
  {Soffitta} P.,  2002, \aap, 390, 81

\bibitem[\protect\citeauthoryear{{Band} et~al.,}{{Band}
  et~al.}{1993}]{band:93}
{Band} D.,  et~al., 1993, \apj, 413, 281

\bibitem[\protect\citeauthoryear{{Band} \& {Preece}}{{Band} \&
  {Preece}}{2005}]{band:05}
{Band} D.~L.,  {Preece} R.~D.,  2005, \apj, 627, 319

\bibitem[\protect\citeauthoryear{{Barraud}, {Daigne}, {Mochkovitch} \&
  {Atteia}}{{Barraud} et~al.}{2005}]{barraud:05}
{Barraud} C.,  {Daigne} F.,  {Mochkovitch} R.,    {Atteia} J.~L.,  2005, \aap,
  440, 809

\bibitem[\protect\citeauthoryear{{Daigne} \& {Mochkovitch}}{{Daigne} \&
  {Mochkovitch}}{1998}]{daigne:98}
{Daigne} F.,  {Mochkovitch} R.,  1998, \mnras, 296, 275

\bibitem[\protect\citeauthoryear{{Daigne} \& {Mochkovitch}}{{Daigne} \&
  {Mochkovitch}}{2003}]{daigne:03}
{Daigne} F.,  {Mochkovitch} R.,  2003, \mnras, 342, 587

\bibitem[\protect\citeauthoryear{{Daigne} \& {Mochkovitch}}{{Daigne} \&
  {Mochkovitch}}{2007}]{daigne:07}
{Daigne} F.,  {Mochkovitch} R.,  2007, \aap, 465, 1

\bibitem[\protect\citeauthoryear{{Daigne}, {Rossi} \& {Mochkovitch}}{{Daigne}
  et~al.}{2006}]{daigne:06}
{Daigne} F.,  {Rossi} E.~M.,    {Mochkovitch} R.,  2006, \mnras, 372, 1034

\bibitem[\protect\citeauthoryear{{Fenimore} \& {Ramirez-Ruiz}}{{Fenimore} \&
  {Ramirez-Ruiz}}{2000}]{fenimore:00}
{Fenimore} E.~E.,  {Ramirez-Ruiz} E.,  2000, astro-ph/0004176

\bibitem[\protect\citeauthoryear{{Firmani}, {Avila-Reese}, {Ghisellini} \&
  {Tutukov}}{{Firmani} et~al.}{2004}]{firmani:04}
{Firmani} C.,  {Avila-Reese} V.,  {Ghisellini} G.,    {Tutukov} A.~V.,  2004,
  \apj, 611, 1033

\bibitem[\protect\citeauthoryear{{Frail}, {Kulkarni}, {Sari}, {Djorgovski},
  {Bloom}, {Galama}, {Reichart}, {Berger}, {Harrison}, {Price}, {Yost},
  {Diercks}, {Goodrich} \& {Chaffee}}{{Frail} et~al.}{2001}]{frail:01}
{Frail} D.~A.,  {Kulkarni} S.~R.,  {Sari} R.,  {Djorgovski} S.~G.,  {Bloom}
  J.~S.,  {Galama} T.~J.,  {Reichart} D.~E.,  {Berger} E.,  {Harrison} F.~A.,
  {Price} P.~A.,  {Yost} S.~A.,  {Diercks} A.,  {Goodrich} R.~W.,    {Chaffee}
  F.,  2001, \apjl, 562, L55

\bibitem[\protect\citeauthoryear{{Ghirlanda}, {Ghisellini}, {Firmani},
  {Celotti} \& {Bosnjak}}{{Ghirlanda} et~al.}{2005}]{ghirlanda:05}
{Ghirlanda} G.,  {Ghisellini} G.,  {Firmani} C.,  {Celotti} A.,    {Bosnjak}
  Z.,  2005, \mnras, 360, L45

\bibitem[\protect\citeauthoryear{{Guetta} \& {Piran}}{{Guetta} \&
  {Piran}}{2007}]{guetta:07}
{Guetta} D.,  {Piran} T.,  2007, Journal of Cosmology and Astro-Particle
  Physics, 7, 3

\bibitem[\protect\citeauthoryear{{Guetta}, {Piran} \& {Waxman}}{{Guetta}
  et~al.}{2005}]{guetta:05}
{Guetta} D.,  {Piran} T.,    {Waxman} E.,  2005, \apj, 619, 412

\bibitem[\protect\citeauthoryear{{Kistler}, {Yuksel}, {Beacom} \&
  {Stanek}}{{Kistler} et~al.}{2007}]{kistler:07}
{Kistler} M.~D.,  {Yuksel} H.,  {Beacom} J.~F.,    {Stanek} K.~Z.,  2007,
  astroph/0709.0381, 709

\bibitem[\protect\citeauthoryear{{Le} \& {Dermer}}{{Le} \&
  {Dermer}}{2007}]{le:07}
{Le} T.,  {Dermer} C.~D.,  2007, \apj, 661, 394

\bibitem[\protect\citeauthoryear{{Liang}, {Zhang}, {Virgili} \& {Dai}}{{Liang}
  et~al.}{2007}]{liang:06}
{Liang} E.,  {Zhang} B.,  {Virgili} F.,    {Dai} Z.~G.,  2007, \apj, 662, 1111

\bibitem[\protect\citeauthoryear{{Nakar} \& {Piran}}{{Nakar} \&
  {Piran}}{2005}]{nakar:05}
{Nakar} E.,  {Piran} T.,  2005, \mnras, 360, L73

\bibitem[\protect\citeauthoryear{{Porciani} \& {Madau}}{{Porciani} \&
  {Madau}}{2001}]{porciani:01}
{Porciani} C.,  {Madau} P.,  2001, \apj, 548, 522

\bibitem[\protect\citeauthoryear{{Preece}, {Briggs}, {Mallozzi}, {Pendleton},
  {Paciesas} \& {Band}}{{Preece} et~al.}{2000}]{preece:00}
{Preece} R.~D.,  {Briggs} M.~S.,  {Mallozzi} R.~S.,  {Pendleton} G.~N.,
  {Paciesas} W.~S.,    {Band} D.~L.,  2000, \apjs, 126, 19

\bibitem[\protect\citeauthoryear{{Rees} \& {Meszaros}}{{Rees} \&
  {Meszaros}}{1994}]{rees:94}
{Rees} M.~J.,  {Meszaros} P.,  1994, \apjl, 430, L93

\bibitem[\protect\citeauthoryear{{Sakamoto} et~al.,}{{Sakamoto}
  et~al.}{2005}]{sakamoto:05}
{Sakamoto} T.,  et~al., 2005, \apj, 629, 311

\bibitem[\protect\citeauthoryear{{Soderberg}, {Kulkarni}, {Berger}, {Fox},
  {Sako}, {Frail}, {Gal-Yam}, {Moon}, {Cenko}, {Yost}, {Phillips}, {Persson},
  {Freedman}, {Wyatt}, {Jayawardhana} \& {Paulson}}{{Soderberg}
  et~al.}{2004}]{soderberg:04}
{Soderberg} A.~M.,  {Kulkarni} S.~R.,  {Berger} E.,  {Fox} D.~W.,  {Sako} M.,
  {Frail} D.~A.,  {Gal-Yam} A.,  {Moon} D.~S.,  {Cenko} S.~B.,  {Yost} S.~A.,
  {Phillips} M.~M.,  {Persson} S.~E.,  {Freedman} W.~L.,  {Wyatt} P.,
  {Jayawardhana} R.,    {Paulson} D.,  2004, \nat, 430, 648

\bibitem[\protect\citeauthoryear{{Stern}, {Atteia} \& {Hurley}}{{Stern}
  et~al.}{2002}]{stern:02a}
{Stern} B.~E.,  {Atteia} J.-L.,    {Hurley} K.,  2002, \apj, 578, 304

\bibitem[\protect\citeauthoryear{{Stern}, {Tikhomirova}, {Kompaneets},
  {Svensson} \& {Poutanen}}{{Stern} et~al.}{2001}]{stern:01}
{Stern} B.~E.,  {Tikhomirova} Y.,  {Kompaneets} D.,  {Svensson} R.,
  {Poutanen} J.,  2001, \apj, 563, 80

\bibitem[\protect\citeauthoryear{{Stern}, {Tikhomirova}, {Stepanov},
  {Kompaneets}, {Berezhnoy} \& {Svensson}}{{Stern} et~al.}{2000}]{stern:00}
{Stern} B.~E.,  {Tikhomirova} Y.,  {Stepanov} M.,  {Kompaneets} D.,
  {Berezhnoy} A.,    {Svensson} R.,  2000, \apjl, 540, L21

\bibitem[\protect\citeauthoryear{{Stern}, {Tikhomirova} \& {Svensson}}{{Stern}
  et~al.}{2002}]{stern:02b}
{Stern} B.~E.,  {Tikhomirova} Y.,    {Svensson} R.,  2002, \apj, 573, 75

\bibitem[\protect\citeauthoryear{{Yamazaki}, {Yonetoku} \&
  {Nakamura}}{{Yamazaki} et~al.}{2003}]{yamazaki:03}
{Yamazaki} R.,  {Yonetoku} D.,    {Nakamura} T.,  2003, \apjl, 594, L79

\bibitem[\protect\citeauthoryear{{Yonetoku}, {Murakami}, {Nakamura},
  {Yamazaki}, {Inoue} \& {Ioka}}{{Yonetoku} et~al.}{2004}]{yonetoku:04}
{Yonetoku} D.,  {Murakami} T.,  {Nakamura} T.,  {Yamazaki} R.,  {Inoue} A.~K.,
    {Ioka} K.,  2004, \apj, 609, 935

\end{thebibliography}

\label{lastpage}
\end{document}